\documentclass[twocolumn]{revtex4}
\usepackage{graphicx}
\usepackage{amssymb}

\makeatletter

\providecommand{\tabularnewline}{\\}

\newcommand{\dblline}[2]{\\ \hline\hline}

\makeatother
\begin{document}

\title{A Study on the Noise Threshold of Fault-tolerant Quantum Error Correction}

\author{Y. C. Cheng }

\author{R. J. Silbey}

\email{silbey@mit.edu}

\affiliation{Department of Chemistry and Center for Materials Science and Engineering\\
Massachusetts Institute of Technology\\
Cambridge, Massachusetts 02139}

\begin{abstract}
Quantum circuits implementing fault-tolerant quantum error correction
(QEC) for the three qubit bit-flip code and five-qubit code are studied.
To describe the effect of noise, we apply a model based on a generalized
effective Hamiltonian where the system-environment interactions are
taken into account by including stochastic fluctuating terms in the
system Hamiltonian. This noise model enables us to investigate the
effect of noise in quantum circuits under realistic device conditions
and avoid strong assumptions such as maximal parallelism and weak
storage errors. Noise thresholds of the QEC codes are calculated.
In addition, the effects of imprecision in projective measurements,
collective bath, fault-tolerant repetition protocols, and level of
parallelism in circuit constructions on the threshold values are also
studied with emphasis on determining the optimal design for the fault-tolerant
QEC circuit. These results provide insights into the fault-tolerant
QEC process as well as useful information for designing the optimal
fault-tolerant QEC circuit for particular physical implementation
of quantum computer.
\end{abstract}
\maketitle

\section{Introduction}

Recent developments in the theory of quantum computation have generated
significant interest in utilizing quantum mechanics to achieve new
computational capability \cite{nielsen_chuang}. A quantum computer
can outperform its classical counterpart and provide efficient ways
to solve many important problems. However, the intrinsic sensitivity
of a quantum superposition state to imperfect operations and interactions
with its surrounding environment prohibits the realization of a scalable
quantum computer. To combat the inevitable errors and decoherence
of quantum states during the process of computation, quantum error
correction (QEC) and fault-tolerant methods of quantum computation
have to be applied in the construction of large-scale quantum computers.
It has become clear that the future of robustly storing and manipulating
quantum information rely upon the success of fault-tolerant quantum
error correction \cite{PWS96,DS96,Got98}.

Fault-tolerant methods combined with concatenated coding yield the
threshold result, that states if the noise level per elementary operation
is below a threshold value, then arbitrarily long quantum computation
can be achieved using faulty components \cite{zalka1996,aharonov:stoc1997,knill:prsla1998,preskill:prsla1998}.
Using a $t$-error correcting code, fault-tolerant circuits constructed
from faulty gates with error rate $\epsilon$ can achieve a logical
error rate of $O(\epsilon^{t+1})$ per logical gate. This fact together
with the concept of concatenated coding provide a method for possible
large-scale quantum computation, and can lead to the realization of
a scalable quantum computer. Therefore, it is important to study fault-tolerant
methods and estimate the noise threshold values. In addition, the
noise threshold indicates the tolerable noise level in a certain quantum
circuit, and provides a benchmark for the efficiency of QEC circuits.

A number of theoretical estimates of noise threshold and improvements
for the efficiency of QEC circuits have been proposed \cite{zalka1996,aharonov:stoc1997,knill:prsla1998,preskill:prsla1998,Ste99,Ste03,Rei04};
however, these studies adopt ad-hoc classical stochastic noise models
that neglect device details, and make strong assumptions such as maximal
parallelism and low noise level in storing qubits. Realistically,
these assumptions are not usually applicable, and the power of fault-tolerant
QEC under realistic physical conditions is still unclear. Noise threshold
values are of little use if limitations of the physical implementation
and realistic noise sources are not considered in the estimation.
Therefore, it is of importance to study fault-tolerant QEC circuits
using a noise model that reflects realistic device conditions.

In Ref. \cite{cheng_pra2004}, we applied a phenomenological noise
model to study the effect of noise in quantum teleportation and controlled-NOT
gate operation. Starting from a effective system Hamiltonian that
incorporates stochastic fluctuating terms to describe the effect of
system-environment interactions, this model can describe the dissipative
dynamics of a many-qubit system under realistic device conditions.
In this paper, the same model is applied to investigate the performance
of fault-tolerant QEC circuits implementing three qubit bit-flip code
and five-qubit code. Relatively small codes are studied because we
perform a systematic investigation on several variables that can affect
the performance of fault-tolerant QEC circuits. In section \ref{sec:Interactions-and-Noise}
we first present the model Hamiltonian we used to implement quantum
gates, and briefly review the noise model we proposed. We then introduce
the fault-tolerant QEC circuits studied in this work in section \ref{sec:Fault-Tolerant-QEC-Circuit},
and show our estimates of noise threshold in section \ref{sec:Estimate-of-Noise}.
Finally, a systematic study on how factors like imperfect measurement,
collective bath, repetition protocol, and level of parallelism affect
the performance of fault-tolerant QEC is presented in section \ref{sec:Efficiency-of-F-T}.
This theoretical study will be useful for the design and implementation
of fault-tolerant QEC circuits. We briefly conclude our results in
section \ref{sec:Conclusion}.

\section{Interactions and Noise Model\label{sec:Interactions-and-Noise}}

We study the performance of fault-tolerant QEC circuits using a microscopic
model described in Ref. \cite{cheng_pra2004}. In this model, a qubit
system is described by a Hamiltonian with a controlled part and a
time dependent stochastic part. The general Hamiltonian of the qubit
system can be written as ($\hbar=1$)

\begin{equation}
\begin{array}{rcl}
\mathbf{H}(t) & = & \mathbf{H}_{0}(t)+\mathbf{h}(t),\end{array}\label{eq:hamiltonian}\end{equation}
 where the controlled Hamiltonian $\textbf{H}_{0}(t)$ describes the
interactions between qubits, and the stochastic part $\textbf{h}(t)$
describes the fluctuations of the interactions due to the coupling
to the environment. During the process of quantum computation, $\textbf{H}_{0}(t)$
is controlled to implement gate operations, whereas $\textbf{h}(t)$
is stochastic and results in the decoherence of the quantum system.

\begin{figure}
\begin{center}\includegraphics[%
  width=0.90\columnwidth,
  keepaspectratio]{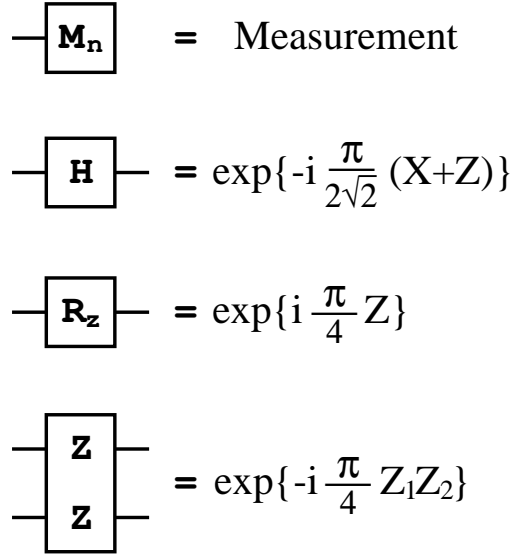}\end{center}

\caption{\label{fig:Quantum-gate-symbols} Quantum gate symbols used to denote
unitary transformations implemented with single-qubit $X$, $Z$ and
two-qubit $ZZ$ interactions. More complicated transformations such
as controlled-NOT and controlled-Z gates can be trivially constructed
using these elementary gates.}
\end{figure}

We choose to simulate fault-tolerant QEC circuits using a model control
Hamiltonian with single-qubit $X$, $Z$ and two-qubit $ZZ$ interactions:

\begin{equation}
\mathbf{H}_{0}(t)={\displaystyle \sum_{i}\varepsilon_{i}(t)Z_{i}+}{\displaystyle \sum_{i}J_{i}(t)X_{i}}+{\displaystyle \sum_{i,j<i}g_{ij}(t)Z_{i}Z_{j},}\label{eq:controll_H}\end{equation}
where $Z_{i}$ and $X_{i}$ are Pauli operators acting on the $i$-th
qubit, and $\varepsilon_{i}(t)$, $J_{i}(t)$, and $g_{ij}(t)$ are
controllable parameters that can be turn on and off to implement desired
gate operations. For simplicity, all gate operations are simulated
using step function pulses with field strengths set to 1 (uniform
field strengths), and the {}``on-time'' of each pulses are controlled
to obtain the desired unitary transformations. Note that by doing
so we adopt a dimensionless system in which a unit time scale $\Delta t$
is defined by the field strength $\varepsilon$, i.e. $\Delta t=1/\varepsilon$.
We consider fault-tolerant QEC circuits composed only of single-qubit
Hadamard gates, two-qubit controlled-NOT and controlled-Z gates, plus
measurement of a single qubit in the computational basis. All these
operations can be easily implemented using the model Hamiltonian in
Eq. (\ref{eq:controll_H}). Figure \ref{fig:Quantum-gate-symbols}
shows the gate symbols and corresponding unitary transformations used
in our simulations. Note that we adopt the $ZZ$ type two-qubit coupling
in our model Hamiltonian for an illustrative purpose. The real form
of the inter-qubit interaction depends on the controllable interactions
available for each individual physical implementations. Nevertheless,
our model can handle the other types of interactions as well, and
we expect that the model Hamiltonian we use here can reproduce the
same general physical behavior as other two-qubit Hamiltonians. 

To describe the effect of noise, we consider fluctuations on the system
Hamiltonian due to system-environment interactions: 

\[
\mathbf{h}(t)={\displaystyle \sum_{i}\delta\varepsilon_{i}(t)Z_{i}+}{\displaystyle \sum_{i}\delta J_{i}(t)X_{i}},\]
where $\delta\varepsilon_{i}(t)$ and $\delta J_{i}(t)$ describe
the time-dependent diagonal and off-diagonal fluctuations on the $i$-th
qubit, respectively. This corresponds to stochastic single-qubit phase
($Z)$ and bit-flip ($X$) errors on each individual qubit. In addition,
we consider the fluctuations as random Gaussian Markov processes with
zero mean and $\delta$-function correlation times described by the
following set of equations:

\begin{equation}
\begin{array}{rcl}
\langle\delta\varepsilon_{i}(t)\rangle & = & \langle\delta J_{i}(t)\rangle=0,\\
\langle\delta\varepsilon_{i}(t)\delta\varepsilon_{j}(t')\rangle & = & \gamma_{0}\cdot\delta_{ij}\delta(t-t'),\\
\langle\delta J_{i}(t)\delta J_{j}(t')\rangle & = & \gamma_{1}\cdot\delta_{ij}\delta(t-t'),\\
\langle\delta\varepsilon_{i}(t)\delta J_{j}(t')\rangle & = & 0,\end{array}\label{eq:stochastic}\end{equation}
where bracket $\langle\rangle$ means averaging over the stochastic
variables, and $\gamma_{0}$ and $\gamma_{1}$ describe the strength
of the diagonal energy fluctuations and off-diagonal matrix element
fluctuations, respectively. For a single-qubit system, $\gamma_{0}$
and $\gamma_{1}$ are well-defined physical quantities, i.e. $\gamma_{0}$
and $\gamma_{1}$ are population relaxation rate and pure dephasing
rate, respectively \cite{cheng_pra2004}. Note that noise strengths
$\gamma_{0}$ and $\gamma_{1}$ should be interpreted as the error
rate per unit time scale $\Delta t=1/\varepsilon$, where $\varepsilon$
is the strength of the control fields. Also notice that we treat the
correlation between different qubits independently, which means each
qubit in the system is coupled to a distinct environment (bath). Later
we will remove this constraint and examine the effect of a collective
bath on the noise threshold value. We also assume that the diagonal
and off-diagonal fluctuations are not correlated. 

For simplicity, we assume that the noise strengths are uniform, i.e.
$\gamma_{0}$ and $\gamma_{1}$ are constants. The noise strength
is set to be the same on all qubits at all times, therefore, we do
not distinguish storage and gate errors. By assuming that the storage
and gate errors are at the same level, the uniform noise assumption
overestimates the errors in the system. At the same time it also avoids
the weak storage noise assumption usually made in previous estimates
of noise thresholds. Realistically, to perform a quantum gate between
two distant qubits in a large-scale quantum circuit, multiple quantum
swap gates must be employed to shuffle quantum states around \cite{FHH04}.
Our uniform noise assumption reflects the physical condition in this
scenario. Note that more complex setups, in which control field and
noise strengths are different for each individual qubits can be studied
with exactly the same method.

Using the method described in Ref. \cite{cheng_pra2004} and Eq. (\ref{eq:hamiltonian})-(\ref{eq:stochastic}),
we can derive the equation of motion for the density matrix of the
qubit system, and numerically propagate the density matrix of a system
with up to twelve qubits (bound by the size of physical memory on
a personal computer). This method provides an efficiently way to simulate
quantum circuits and obtain full dynamics of the qubit system.

\section{Fault-Tolerant QEC Circuit\label{sec:Fault-Tolerant-QEC-Circuit}}

In this paper, we study fault-tolerant QEC circuits implementing the
three qubit bit-flip code and five qubit code. We choose to investigate
these two codes, because they are relatively small and allow us to
perform systematic studies. Previous studies on the fault-tolerant
QEC have been mainly focused on CSS codes, especially the CSS {[}{[}7,1,3{]}{]}
code \cite{CS96,Ste96,Ste99}. Because fault-tolerant encoded operations
on CSS codes are easy to implement, CSS codes are expected to be more
useful for quantum computation than the three qubit bit-flip code
and five qubit code. Nevertheless, since we focus on variables affecting
the performance of fault-tolerant QEC circuits, we expect that results
gained in our study can be applied to more general codes. In this
section, we introduce these two codes and the methods we apply to
perform fault-tolerant QEC.

\subsection{Fault-tolerant QEC scheme}

We adopt the fault-tolerant QEC scheme proposed by DiVincenzo and
Shor \cite{DS96}. This protocol utilizes cat states and transversal
controlled-$X/Z$ gates to detect error syndromes. The procedure can
be divided into three different stage: (1) ancilla preparation and
verification, (2) syndrome detection, and (3) recovery.

\begin{figure}
\begin{center}\includegraphics[%
  width=0.90\columnwidth,
  keepaspectratio]{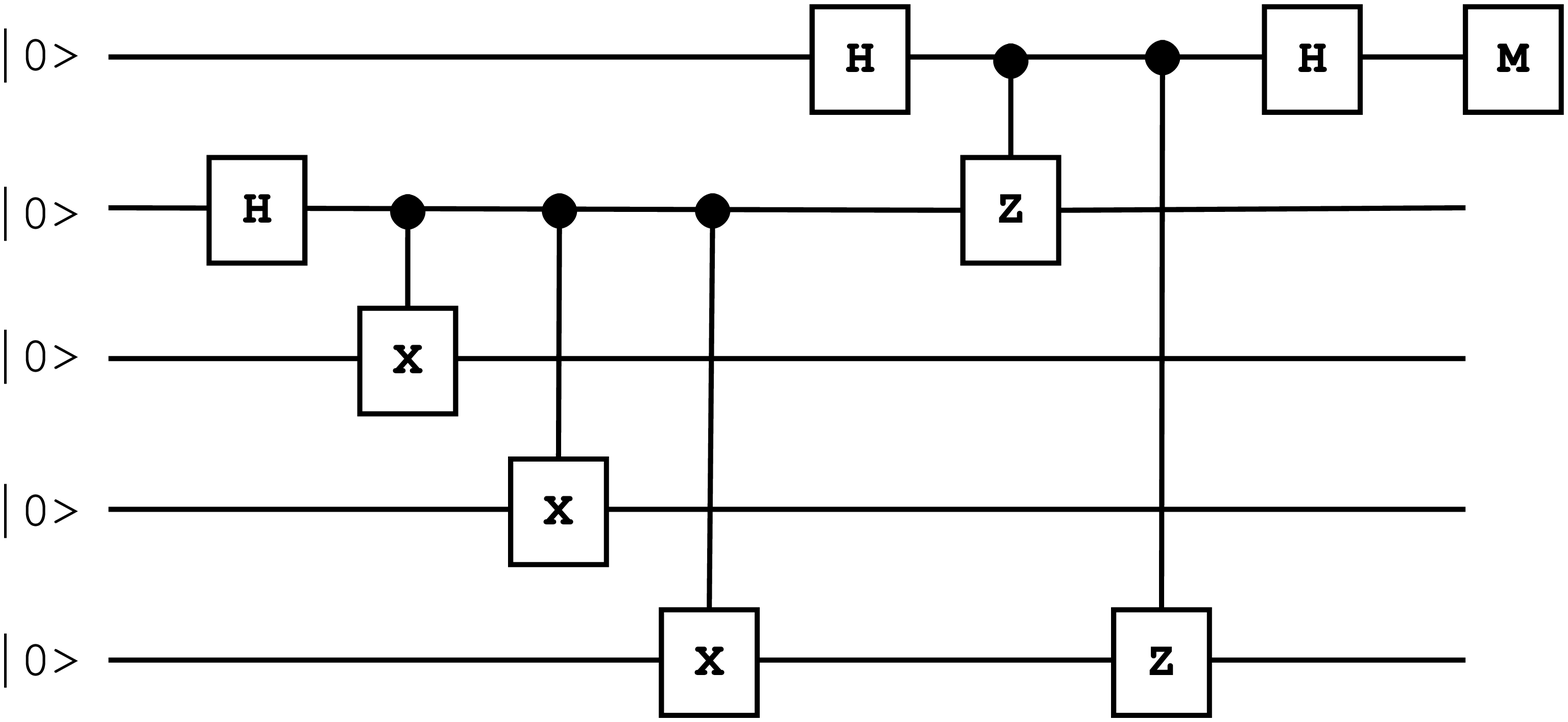}\end{center}

\caption{\label{fig:4ghz_preparation} The fault-tolerant circuit for the
preparation and verification of the four-qubit cat state. Note that
the final result is conditioned by the outcome of the measurement
at the end of the circuit. If the measurement outcome is zero, we
accept the state; otherwise, the state is discarded and the circuit
is started over again.}
\end{figure}

To detect syndrome fault-tolerantly, ancilla qubits have to be prepared
in maximally entangled cat states, and go through a verification step
to ensure that magnitudes of correlated multiple-qubit errors are
small. For example, the four-qubit cat state $\frac{1}{\sqrt{2}}(|0000\rangle+|1111\rangle)$
is necessary for the fault-tolerant QEC of the five-qubit code. Figure
\ref{fig:4ghz_preparation} shows the circuit we used to prepare and
verify four-qubit cat states \cite{preskill:prsla1998}. In this circuit,
an extra qubit is used to detect correlated $X$ errors in the cat
state; after the measurement, only states with measurement result
equals to zero are accepted. This verification step ensures that a
single-qubit error in the circuit causes at most a single-qubit error
in the final cat state, therefore the circuit fulfills the fault-tolerant
condition. Compared to other fault-tolerant cat state preparation
circuits \cite{nielsen_chuang,steane0202036}, an important feature
in the circuit in Fig. \ref{fig:4ghz_preparation} is that only a
projective measurement is required to verify the cat state fault-tolerantly.
This is possible because the circuit takes into account the error
propagation pattern in the preparation step. 

The ancilla cat state generated by the circuit presented in Fig. \ref{fig:4ghz_preparation}
is used to perform transversal controlled-stabilizer operation to
transfer information about the errors from the data qubits to the
ancilla qubits. After decoding the ancilla state, projective measurement
is then applied to obtain error syndromes. Because there are more
gates in the circuits than the number of measurements, it is reasonable
to assume that measurement has smaller effect on the threshold result.
Therefore, we assume perfect measurement. Later we will study the
effect of measurement errors. In addition, to ensure that we do not
accept a wrong syndrome, we must repeat syndrome detection and take
a majority vote. Following Shor's protocol, the following repetition
scheme is used:

\begin{description}
\item [Repetition~Protocol~A~(three~majority~vote):]~
\end{description}
\begin{enumerate}
\item Perform the syndrome detection twice. If the same measurement results
are obtained, the syndrome is accepted and data qubits are corrected.
\item Otherwise, perform one more syndrome detection. If any two of the
three measurement results are the same, the syndrome is accepted and
data qubits are corrected.
\item If all three measurement results are different, no further action
is taken.
\end{enumerate}
This protocol is basically a simple majority vote in three trials.
Note that the choice of the repetition protocol is not unique. In
fact, later we will compare protocol A to another protocol, and show
that we can improve this protocol to increase the efficiency of the
fault-tolerant QEC procedure.

Combining all these elements, we can ensure that the probability of
generating a two-qubit error is of order $\epsilon^{2}$, and avoid
the catastrophic propagation of errors. During a quantum computation
using a single-error correcting code, we can perform the fault-tolerant
QEC after each gate operations. In consequence, single-qubit errors
generated in earlier computation and QEC steps will be corrected in
later QEC steps. Therefore, only two-qubit errors will be accumulated
in a rate of order $\epsilon^{2}$. As a result, we can achieve longer
computation when $\epsilon$ is small.

\subsection{Three qubit bit-flip code}

The three qubit bit-flip code encodes a logical qubit in three physical
qubits using the following logical states:

\[
\begin{array}{rcl}
|0_{L}\rangle & = & |000\rangle,\\
|1_{L}\rangle & = & |111\rangle.\end{array}\]
This code is a stabilizer code with two stabilizer operators $g_{1}=ZZI$
and $g_{2}=IZZ$. The three qubit bit-flip code corrects single bit-flip
error on any of the three data qubits. This code does not correct
phase errors, therefore it is only useful when the degradation of
the quantum state is dominated by bit-flip errors. However, we believe
insights gained by studying this code can be applied to more general
quantum error-correcting codes. 

\begin{figure}
\begin{center}\includegraphics[%
  width=0.90\columnwidth,
  keepaspectratio]{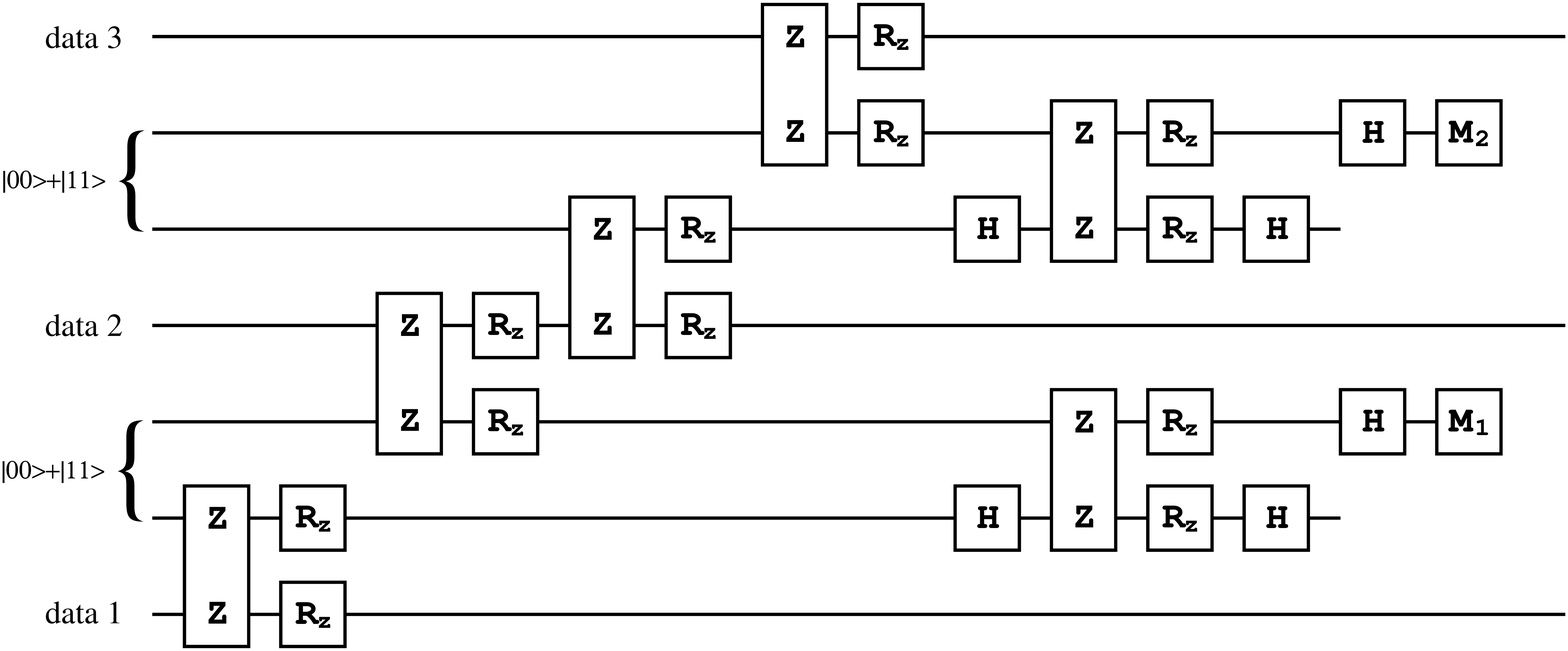}\end{center}

\caption{\label{fig:3qbfc_symdrome_detection_implementation} A circuit implementing
the fault-tolerant syndrome detection for the three qubit bit-flip
code.}
\end{figure}

Figure \ref{fig:3qbfc_symdrome_detection_implementation} shows the
syndrome detection circuit for the three qubit bit-flip code. In this
circuit, physical qubits are depicted by horizontal solid lines, and
quantum gates are represented by boxes. The quantum circuit includes
three data qubits that take a encoded state as the input, and two
pairs of ancilla qubits prepared in the Bell state $\frac{1}{\sqrt{2}}(|00\rangle+|11\rangle)$,
which are used to measure the syndromes. We want to point out that
only limited ability to perform operations in parallel is assumed
in constructing this circuit. In addition, the circuit is arranged
to minimize error propagation from the ancilla qubits to the data
qubits. At the end of the circuit, two measurements, $M_{1}$ and
$M_{2}$, are performed to obtain the error syndrome. After the syndrome
is confirmed according to the repetition protocol A, we then apply
the corresponding recovery action to correct the detected error. Table
\ref{tab:syndrome_3qbfc} lists the syndrome and the corresponding
recovery actions for the three qubit bit-flip code. 

Because the three qubit bit-flip code only corrects bit-flip errors,
we only consider off-diagonal fluctuations on each qubit when dealing
with this code ($\gamma_{0}=0$). Note that the circuit does not protect
against $Z$ errors, nor can it prevent the generation of $Z$ errors.
To access its performance on controlling $X$ errors on the data qubits,
we only study the fault-tolerant QEC procedure when initially the
data qubits are in the logical $|0_{L}\rangle$ state. This selection
of initial state is unrealistic, but it allows us to avoid uncorrectable
$Z$ errors that will ruin the QEC procedure.

\begin{table}

\caption{\label{tab:syndrome_3qbfc}Measurement results and the corresponding
actions required to correct the error in the data qubit for the three
qubit bit-flip code.}

\begin{center}\begin{tabular}{ccccccccc}
&
&
&
&
&
&
&
&
\tabularnewline
\hline
\hline 
\multicolumn{4}{c}{Syndrome}&
&
&
\multicolumn{3}{c}{Action}\tabularnewline
\cline{1-4} \cline{7-9} 
&
M$_{1}$&
M$_{2}$&
&
&
&
&
U$_{R}$&
\tabularnewline
\hline
&
0&
0&
&
&
&
&
III&
\tabularnewline
&
0&
1&
&
&
&
&
IIX&
\tabularnewline
&
1&
0&
&
&
&
&
XII&
\tabularnewline
&
1&
1&
&
&
&
&
IXI&
\tabularnewline
\hline
\hline 
&
&
&
&
&
&
&
&
\tabularnewline
\end{tabular}\end{center}
\end{table}

\subsection{The five-qubit code}

The five-qubit code is the smallest quantum code that corrects all
single-qubit errors \cite{bennett:mixed_state,LMP+96}. A scheme for
fault-tolerant quantum computation using five-qubit code is presented
by Gottesman in Ref. \cite{Got98}. Here we adopt the representation
and fault-tolerant QEC circuit presented by DiVincenzo and Shor in
Ref. \cite{DS96}. Their implementation uses a nine-qubit system with
five data qubits and four ancilla qubits, which utilizes four-qubit
cat state $\frac{1}{\sqrt{2}}(|0000\rangle+|1111\rangle)$ for syndrome
detection. Moreover, four syndromes are detected sequentially. It
is straightforward to simulate the syndrome detection circuit presented
in their paper using our choice of model Hamiltonian (Eq. \ref{eq:hamiltonian}-\ref{eq:stochastic}).

Ideally, multiple input states have to be studied to obtain averaged
performance of the QEC procedure. To avoid such tedious computations,
we use a logical qubit initially in the following pure state density
matrix (in the $\{|0_{L}\rangle,|1_{L}\rangle\}$ basis):

\[
\rho_{0}=\frac{1}{2}(I+\frac{1}{\sqrt{3}}X+\frac{1}{\sqrt{3}}Y+\frac{1}{\sqrt{3}}Z).\]
This state provides an averaged measure for all possible logical states,
thus should give us a reasonable estimate of the averaged circuit
performance. 

Note that our setup simulates a minimal circuit for the fault-tolerant
QEC using five-qubit code with limited physical resources. We expect
such nine-qubit system can be realized on a liquid-state NMR quantum
computer using available technologies. An experimental study on such
minimal fault-tolerant QEC circuit will be an excellent test for our
noise model, and can also provide us invaluable information that is
essential for the design of large-scale quantum computers.

\section{Estimate of Noise Threshold\label{sec:Estimate-of-Noise}}

To estimate the noise threshold for a logical operation, we simulate
a computation where fault-tolerant QEC is performed after each logical
operation on the encoded qubits, and compare the magnitude of logical
errors to the magnitude of errors generated by the same operation
on a bare physical qubit without QEC. We use the crash probability
$P_{c}$ to describe the amount of logical errors in an encoded state.
The crash probability is defined as the probability of having an uncorrectable
error in the data qubits, and can be obtained from the fidelity of
the state \emph{after a perfect QEC process}. 

We define a computational step as a logical gate followed by a fault-tolerant
QEC step. If the same computational step is applied on the data qubits
repeatedly $n$ times, we can describe the crash probability as a
function of $n$, i.e. $P_{c}=P_{c}(n)$. In general, $P_{c}(n)$
satisfies an exponential form:

\begin{equation}
P_{c}(n)=\frac{1}{2}(1-e^{-2\Gamma_{n}n}).\label{eq:Pcn_form}\end{equation}
We can perform simulation and compute crash probability at each step,
$P_{c}(n)$. By fitting our simulation result to the functional form
in Eq. (\ref{eq:Pcn_form}), we obtain the crash rate constant per
computational step $\Gamma_{n}=\left.\frac{dP_{c}(n)}{dn}\right|_{n=0}$.
In addition, we also define the crash rate constant per unit time
$\Gamma_{t}=\left.\frac{dP_{c}(t)}{dt}\right|_{t=0}=\Gamma_{n}/\tau$,
where $\tau$ is the time period required to complete a computational
step. Note that the unit time scale $\Delta t$ is defined by the
strength of control fields $\varepsilon$, $\Delta t=1/\varepsilon$.

\begin{figure}
\begin{center}\includegraphics[%
  width=0.45\columnwidth]{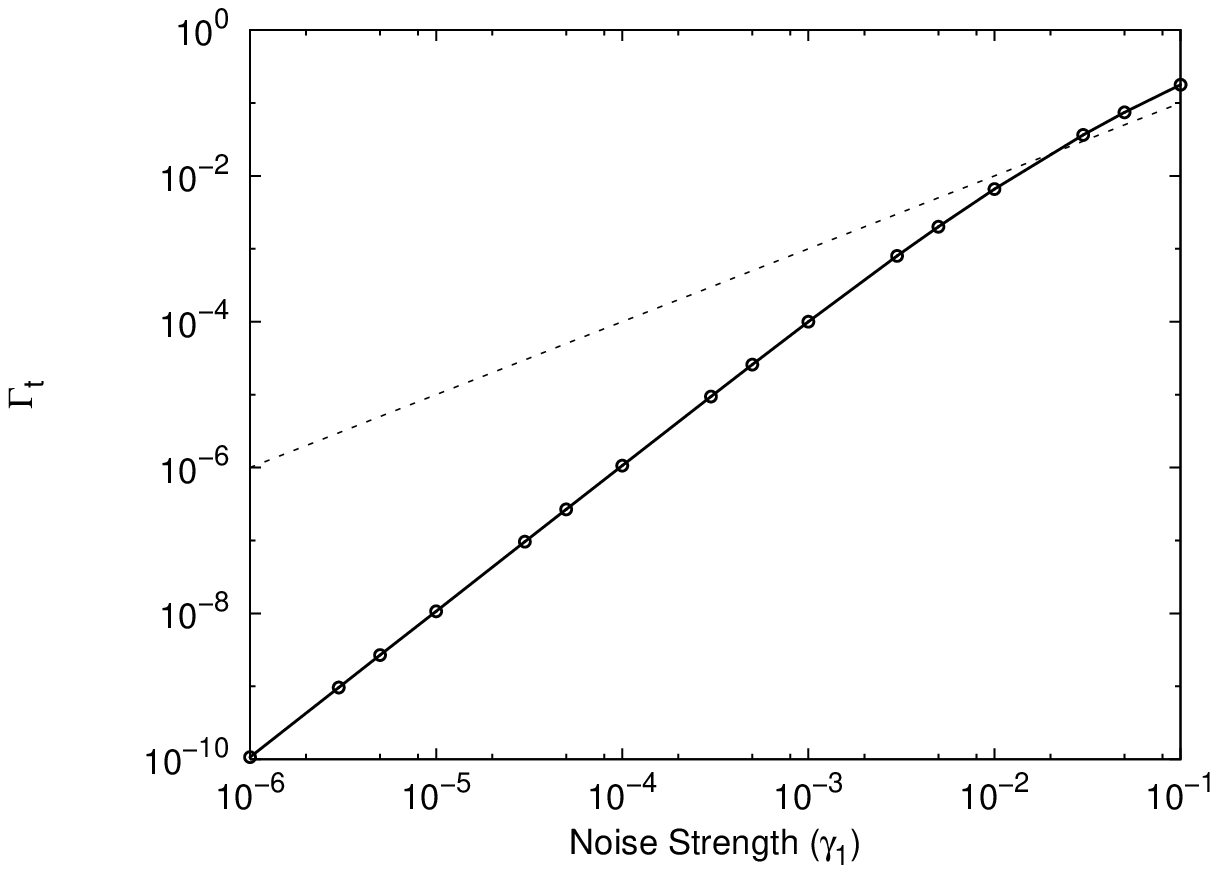}\includegraphics[%
  width=0.45\columnwidth]{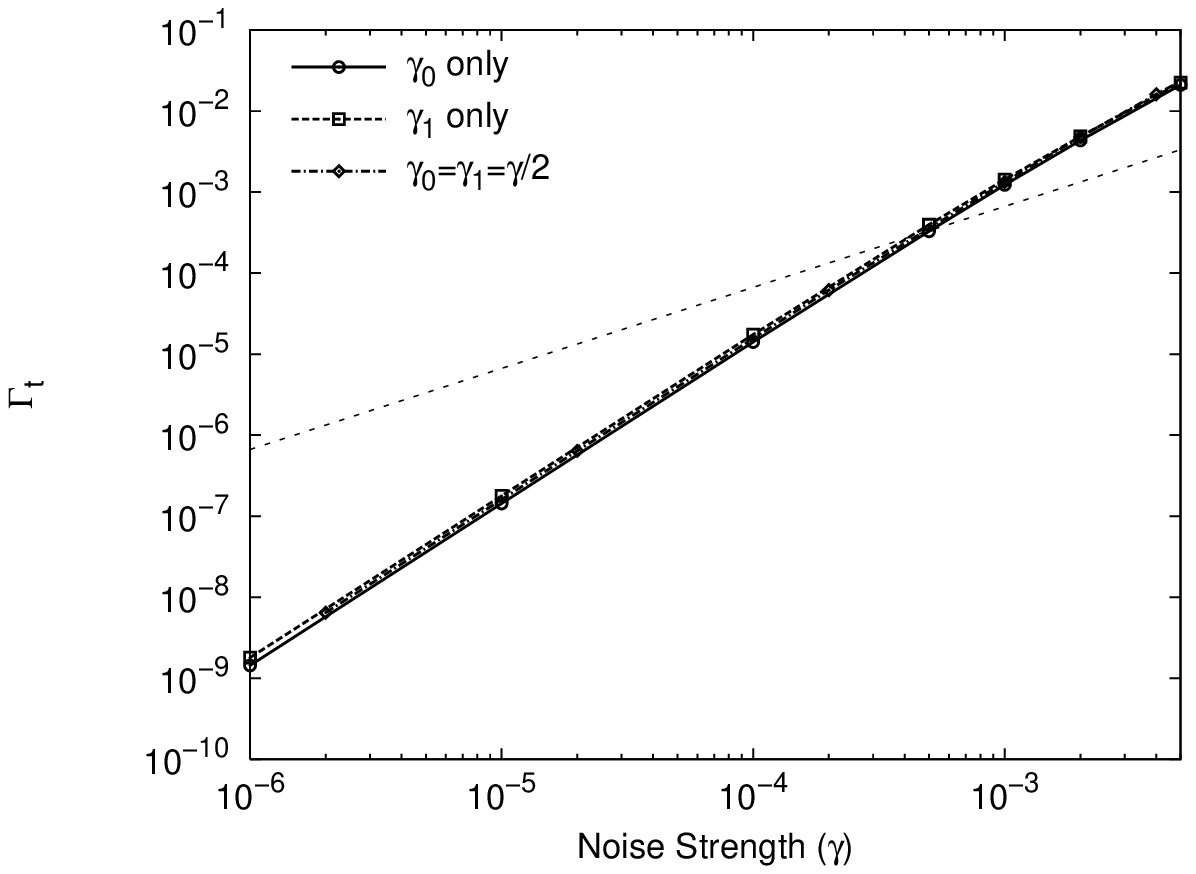}\end{center}

\begin{center}\includegraphics[%
  width=0.45\columnwidth]{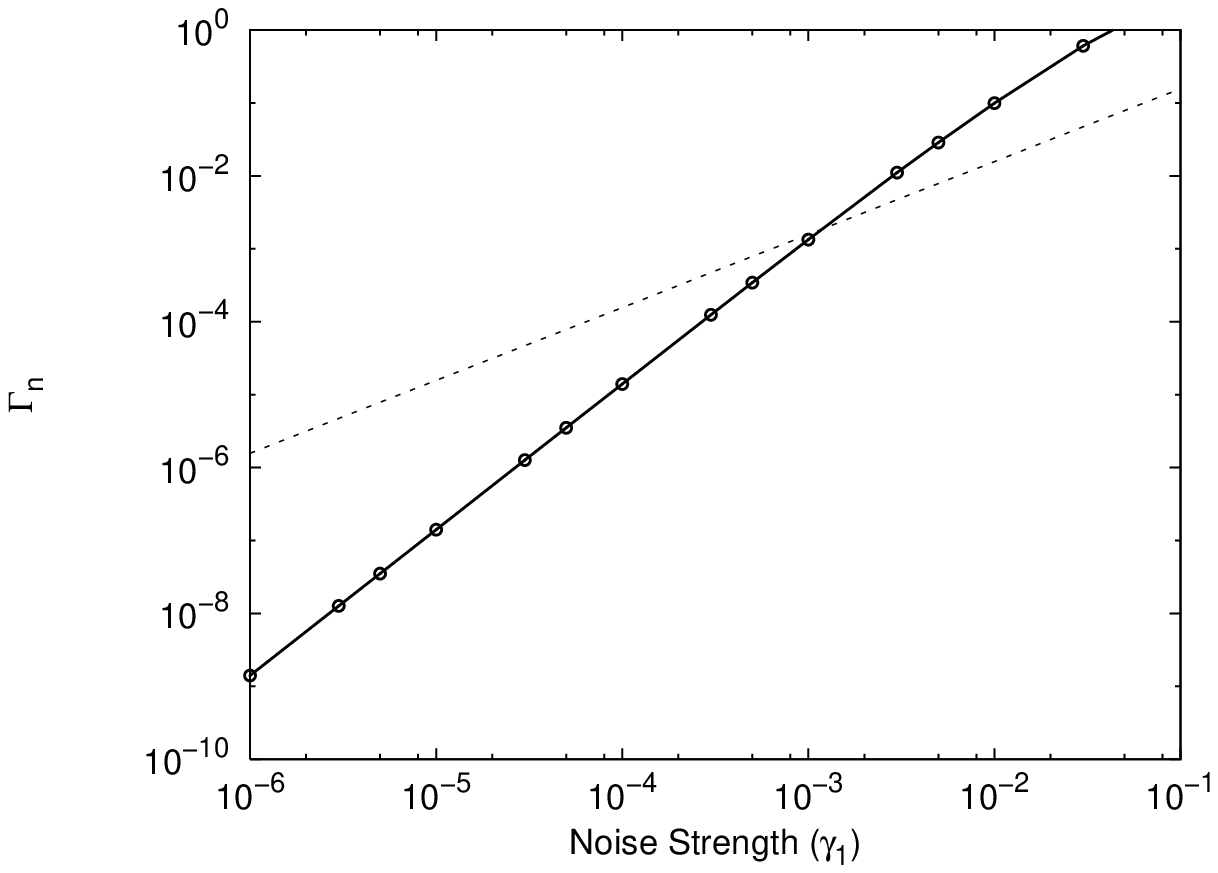}\includegraphics[%
  width=0.45\columnwidth]{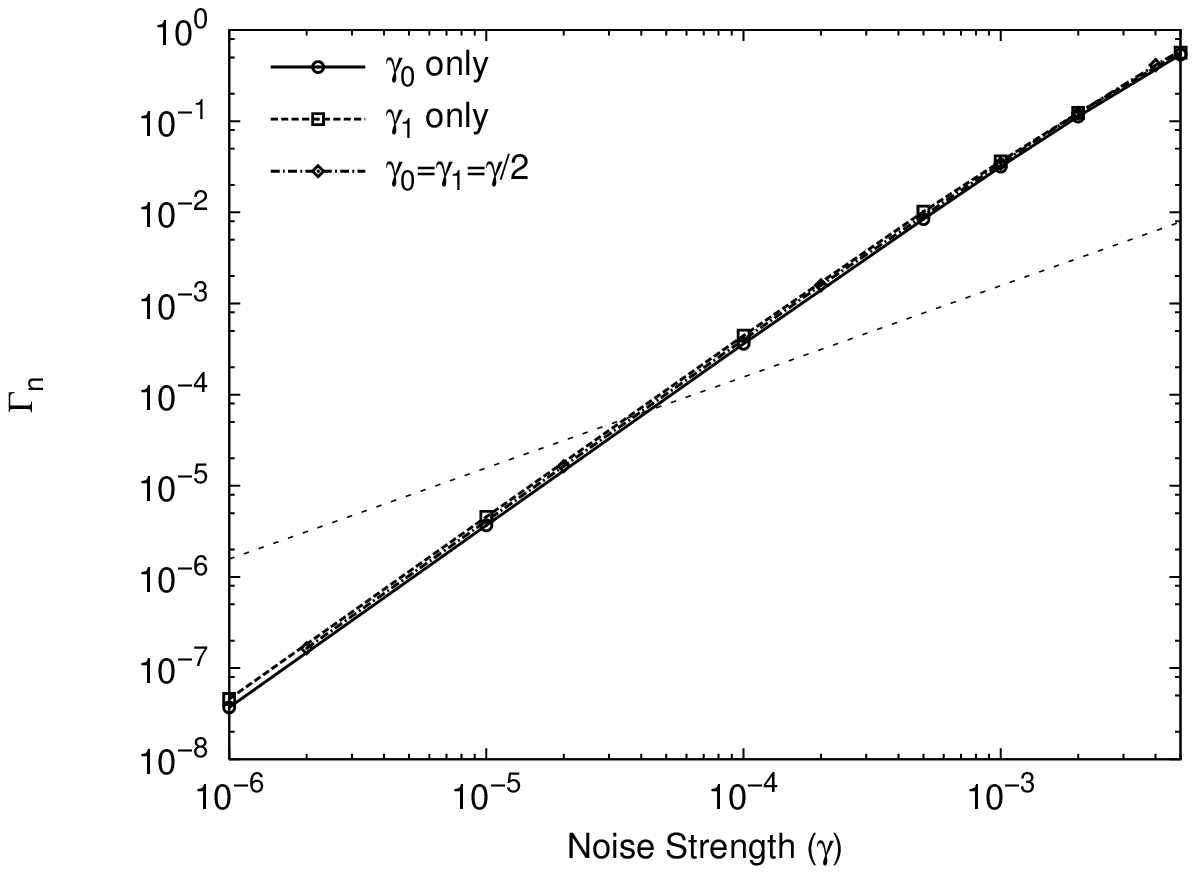}\end{center}

\caption{\label{fig:ftqec_3qbfc} Crash rate constants as a function of the
noise strength. We show crash rate constants for a quantum memory
using the three qubit bit-flip code (upper-left) and five-qubit code
(upper-right), and for a logical $X$ gate on the three qubit bit-flip
code (bottom-left) and on the five-qubit code (bottom-right). For
the five-qubit code circuits, curves for different types of noise
are presented. To show the threshold result, we also present curves
for the error rate of a single physical qubit (dotted line). The noise
threshold values are summarized in Table \ref{tab:Summary-noise-threshold}. }
\end{figure}

We compute noise threshold for a quantum memory, where repeated fault-tolerant
QEC is applied on the data qubits to stabilize quantum information;
and logical $X$ gate, where a logical $X$ gate followed by a fault-tolerant
QEC step are applied on the data qubits. Figure \ref{fig:ftqec_3qbfc}
shows the crash rate constants as a function of noise strength for
the three qubit bit-flip code, as well as the results for the five-qubit
code. In Fig. \ref{fig:ftqec_3qbfc}, we clearly see that in the weak
noise regime, the crash rate constant is proportional to the square
of the noise strength, which reflects the power of the fault-tolerant
QEC procedures. The noise threshold is obtained from the critical
value where the crash rate constant for encoded computation cross
over with the error rate of a bare physical qubit. At noise strength
below the threshold value, the errors in the encoded state accumulated
slower than for the bare physical qubit. At noise strength above the
threshold value, the fault-tolerant QEC provides no benefit. For the
three qubit bit-flip code, the noise threshold is about $2\times10^{-2}$
for quantum memory, and about $1\times10^{-3}$ for the logical $X$
gate. 

We also perform calculations on five-qubit code. The five-qubit code
corrects all single-qubit error, so we can compute the threshold for
different types of noise. Table \ref{tab:Summary-noise-threshold}
summarizes threshold values for three qubit bit-flip code and five-qubit
code. The difference in the noise threshold between quantum memory
and logical $X$ gate is mainly due to the different basis of comparison.
For the quantum memory, we need to compare crash rate constant per
unit time $\Gamma_{t}$ to the decay rate of a free physical qubit;
however, for the X gate, we need to use the crash rate constant per
computational step $\Gamma_{n}$. The extra logical $X$ operation
has little effect on the crash rate per computational step because
the fault-tolerant QEC circuit is much bigger than the circuit for
the logical $X$ gate. This observation suggests that other encoded
single-qubit operations and transversal encoded two-qubit operations
should have similar threshold values.

We summarize the assumptions we made for these calculations: (1) stochastic
$X$ and $Z$ noises, (2) each qubit coupled to a distinct bath, (3)
uniform noise strength, (4) perfect physical $|0\rangle$ states as
initial states, (5) perfect instantaneous projective measurement.
Clearly, the uniform noise assumption that treats gate errors and
storage errors on the same footing is responsible for the relatively
low threshold we obtain for the five-qubit code. In the next section
we will examine some of these assumptions and discuss different factors
that might affect the performance of fault-tolerant QEC.

\begin{table*}

\caption{\label{tab:Summary-noise-threshold} Summary of noise threshold values.
The noise strengths ($\gamma_{0}$ and $\gamma_{1}$) should be interpreted
as the error rate per unit time scale $\Delta t=1/\varepsilon$, where
$\varepsilon$ is the strength of the control fields.}

\begin{center}\begin{tabular}{cccccccc}
&
&
&
&
&
&
&
\tabularnewline
\hline
\hline 
&
&
\multicolumn{2}{c}{three qubit bit-flip code}&
&
\multicolumn{2}{c}{five-qubit code}&
\tabularnewline
\cline{3-4} \cline{6-7} 
&
&
quantum memory&
X gate&
&
quantum memory&
X gate&
\tabularnewline
&
&
&
&
&
&
&
\tabularnewline
$X$ errors ($\gamma_{1}$)&
&
$2.1\times10^{-2}$&
$1.2\times10^{-3}$&
&
$4.2\times10^{-4}$&
$3.5\times10^{-5}$&
\tabularnewline
$Z$ errors ($\gamma_{0}$)&
&
-&
-&
&
$5.1\times10^{-4}$&
$4.3\times10^{-5}$&
\tabularnewline
Both $X$ and $Z$ errors&
&
-&
-&
&
$4.7\times10^{-4}$&
$3.9\times10^{-5}$&
\tabularnewline
&
&
&
&
&
&
&
\tabularnewline
\hline
\hline 
&
&
&
&
&
&
&
\tabularnewline
\end{tabular}\end{center}
\end{table*}

\section{Efficiency of Fault-tolerant QEC Circuits\label{sec:Efficiency-of-F-T}}

In this section, we study several variables that can affect the efficiency
of the fault-tolerant QEC scheme. We perform a systematic investigation
on the performance of quantum memories stabilized by fault-tolerant
QEC and aim to generate a generic picture on how these variables change
the efficiency of fault-tolerant QEC circuits.

\subsection{Effect of imperfect measurement}

\begin{figure}
\begin{center}\includegraphics[%
  width=0.90\columnwidth]{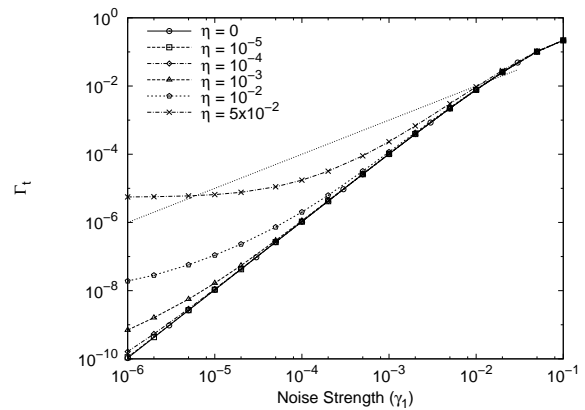}\end{center}

\caption{\label{fig:Pc_measurement} $\Gamma_{t}$ as a function of noise
strength for the fault-tolerant QEC circuit using three qubit bit-flip
code at different level of measurement errors. The error rate of a
single physical qubit is also shown (dotted line). The measurement
error has little effect on the threshold value.}
\end{figure}

We first test the effect of imperfect measurements on the performance
of a quantum memory stabilized using fault-tolerant QEC. We use the
following POVM (positive operator-valued measure) to describe an imperfect
projective measurement on a single qubit:

\[
\begin{array}{rcl}
M_{0} & = & (1-\eta)|0\rangle\langle0|+\eta|1\rangle\langle1|,\\
M_{1} & = & (1-\eta)|1\rangle\langle1|+\eta|0\rangle\langle0|,\end{array}\]
where $M_{0}$ ($M_{1}$) describes events in which basis state $|0\rangle$
($|1\rangle$) is measured, and $\eta$ is the probability of measurement
error, i.e. a projection onto the wrong basis state. Figure \ref{fig:Pc_measurement}
shows curves for the crash rate constant per unit time $\Gamma_{t}$
at different probabilities of measurement errors for a quantum memory
implementing the three qubit bit-flip code. Clearly, $\Gamma_{t}$
is insensitive to the measurement errors even when the probability
of measurement errors is significantly higher than the noise strength
$\gamma_{1}$. The probability of the measurement error as high as
5\% has only minor effect on the threshold value.

\subsection{Effect of a collective bath}

A distinct feature of our noise model is the ability to describe the
effect of a collective bath, in which all qubits are coupled to the
same environment. Such an environment is relevant in physical implementations
such as trapped-ion quantum computers, where qubits are coupled to
the same collective phonon modes \cite{CZ95,WBB+03}. The effect of
a collective bath on the fault-tolerant QEC is an interesting topic.
Because a collective bath seems to contradict the idea of uncorrelated
and stochastic errors that is the foundation of fault-tolerant QEC,
several authors have suggested that collective decoherence has to
be avoided for fault-tolerant quantum computing \cite{preskill:prsla1998,Ste98}.
Also, in a collective bath the effects of noise on different qubits
add coherently; as a result, superdecoherence states exist, and might
affect the efficiency of fault-tolerant QEC \cite{ekert:pmpe1996}.

\begin{figure}
\begin{center}\includegraphics[%
  width=0.90\columnwidth,
  keepaspectratio]{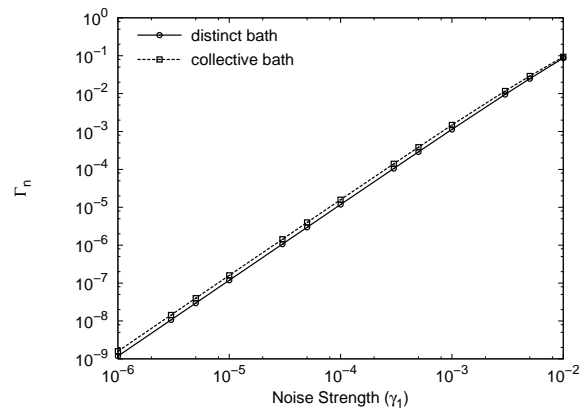}\end{center}

\caption{\label{fig:Pc_bath} The crash rate per computational step $\Gamma_{n}$
for the three qubit bit-flip code as a function of the noise strength.
Curves for the distinct bath system (solid line) and collective bath
system (dotted line) are shown. The result for the collective bath
is close to the result for localized baths. This result suggests that
collective bath has minor effect on the efficiency of fault-tolerant
QEC. }
\end{figure}

To answer this question, we simulate the fault-tolerant QEC circuit
for the three qubit bit-flip code using a noise model in which all
qubits are coupled to a common bath. The following forms of correlation
functions for the stochastic process are used:

\begin{equation}
\begin{array}{rcl}
\langle\delta\varepsilon_{i}(t)\rangle & = & \langle\delta J_{i}(t)\rangle=0,\\
\langle\delta\varepsilon_{i}(t)\delta\varepsilon_{j}(t')\rangle & = & \gamma_{0}\cdot\delta(t-t'),\\
\langle\delta J_{i}(t)\delta J_{j}(t')\rangle & = & \gamma_{1}\cdot\delta(t-t'),\\
\langle\delta\varepsilon_{i}(t)\delta J_{j}(t')\rangle & = & 0,\end{array}\label{eq:sb_stochastic}\end{equation}
Notice that in Eq. (\ref{eq:sb_stochastic}), fluctuations on different
qubits are fully correlated; this reflects the result of coupling
to a common bath. Figure \ref{fig:Pc_bath} shows the crash rate constant
$\Gamma_{n}$ for quantum memories using the three qubit bit-flip
code with two different types of baths. The crash rate curve for the
collective bath case is only slightly higher than the curve for the
localized bath, and there is no significant difference between these
two lines. This result suggests that a collective Markovian bath,
which exhibits spatial but not temporal correlation, has little effect
on the efficiency of fault-tolerant QEC. Although superdecoherence
states do exist when the system is coupled to a collective bath, they
have little effect on the dynamics of the system, because those states
represent only a small fraction in the whole Hilbert space. The fault-tolerant
QEC circuit using the five-qubit code was also studied, and similar
results were obtained. 

\subsection{Repetition protocol}

\begin{figure}
\begin{center}\includegraphics[%
  width=0.90\columnwidth,
  keepaspectratio]{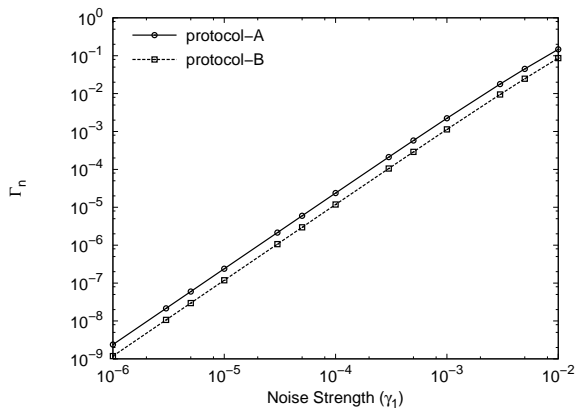}\end{center}

\caption{\label{fig:Pc_protocol} The crash rate constant per step $\Gamma_{n}$
as a function of the noise strength for the two different repetition
protocols for a quantum memory using three qubit bit-flip code. Using
protocol B reduces the crash rate constant by a factor of two.}
\end{figure}

Our simulation propagates the density matrix of the system in the
process of computation, therefore, we obtain the full information
about the time evolution of the system. By examining the trajectory
of the system during the fault-tolerant QEC process, we find the following
repetition protocol yields the best performance: 

\begin{description}
\item [Repetition~Protocol~B~(conditional~generation):]~
\end{description}
\begin{enumerate}
\item Perform the syndrome detection once. If this syndrome is zero, do
nothing.
\item Otherwise, perform the syndrome detection again. If the same syndrome
is obtained, accept the syndrome and correct data qubits accordingly.
\item Otherwise, no further action is taken.
\end{enumerate}
Figure \ref{fig:Pc_protocol} shows the crash rate constant $\Gamma_{n}$
for quantum memories implementing three qubit bit-flip code using
different repetition protocols. Because the majority of the measured
syndromes will be zero in the weak noise regime, protocol B reduces
the amount of time required for a fault-tolerant QEC step by a factor
of two. As a result, the crash rate constant per computational step
$\Gamma_{n}$ decreases by a factor of two when protocol B is used.
Similar improvements on the fault-tolerant QEC protocol have been
suggested by other groups \cite{PVK97,Ste03,Rei04}. The idea behind
protocol B is that the syndrome detection circuit is complicated and
generates extra errors on the data qubits, therefore minimizing the
number of syndrome detection and accepting a syndrome only when two
consecutive detections agree on the same syndrome improve the efficiency
of the fault-tolerant QEC procedure.

\subsection{Level of parallelism}

\begin{figure}
\begin{center}\includegraphics[%
  width=0.90\columnwidth,
  keepaspectratio]{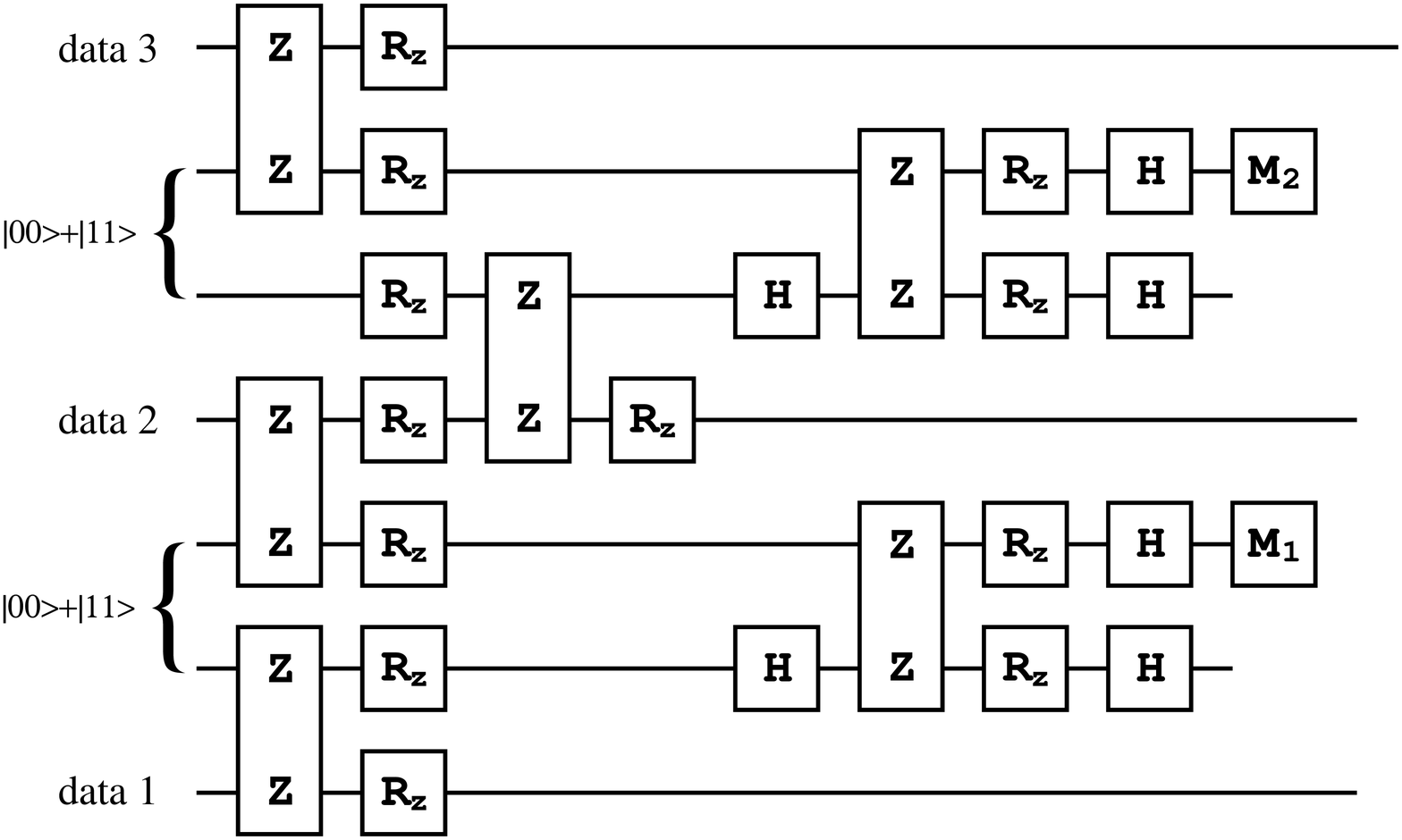}\end{center}

\caption{\label{fig:3qbfc_increased_parallelism} A circuit implementing the
fault-tolerant syndrome detection for the three qubit bit-flipping
code.}
\end{figure}

An important factor related to the efficiency of a QEC circuit is
the level of parallelism in the circuit. The syndrome detection circuit
shown in Fig. \ref{fig:3qbfc_symdrome_detection_implementation} is
not optimized; for example, for a reasonable physical implementation,
the first two controlled-Z gates might actually be operated in parallel
to reduce the operation time. Figure \ref{fig:3qbfc_increased_parallelism}
shows a compressed version of the syndrome detection circuit that
has increased level of parallelism. Furthermore, because the interactions
used to implement the controlled-Z gate commute with each other (Z
and ZZ commute), the controlled-Z gate can be done in one step:

\[
controlled-Z=e^{-i\pi Z_{1}Z_{2}/4}e^{i\pi(Z_{1}+Z_{2})/4}=e^{-i\pi(Z_{1}Z_{2}-Z_{1}-Z_{2})/4}.\]
This makes it possible to perform all controlled-Z operations in a
single pulse. Note that this maximal parallelism design does not generally
exist for arbitrary physical implementations, and is a special case
for our choice of model interactions (ZZ coupling). 

\begin{figure}
\begin{center}\includegraphics[%
  width=0.90\columnwidth,
  keepaspectratio]{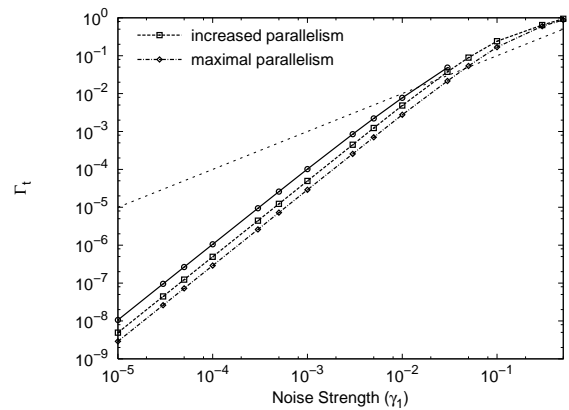}\end{center}

\caption{\label{fig:Pc_parallelism} The crash rate constant per unit time
$\Gamma_{t}$ as a function of the noise strength for quantum memories
using three qubit bit-flip code. Curves for three syndrome detection
circuits different in the level of parallelism are shown. The solid
line is for the circuit shown in Fig. \ref{fig:3qbfc_symdrome_detection_implementation},
the dashed line is the increased parallelism circuit shown in Fig.
\ref{fig:3qbfc_increased_parallelism}, and the dash-dotted line is
the maximal parallelism circuit that finishes all controlled-Z operations
in a single pulse. We see dramatic improvement on the noise threshold
values when the level of parallelism is increased. The result indicates
that by increasing the level of parallelism, the threshold value can
be significantly improved.}
\end{figure}

Figure \ref{fig:Pc_parallelism} shows the crash rate constant per
unit time $\Gamma_{t}$ for quantum memories implementing three qubit
bit-flip code. Results for three syndrome detection circuits with
different level of parallelism are shown. The noise thresholds for
the original circuit (Fig. \ref{fig:3qbfc_symdrome_detection_implementation}),
increased parallelism circuit (\ref{fig:3qbfc_increased_parallelism}),
and maximal parallelism circuit are about $1.5\times10^{-2}$, $2.3\times10^{-2}$,
and $4.6\times10^{-2}$, respectively. The results indicate that by
increasing the level of parallelism, the noise threshold can be significantly
improved. Note that the reduction of the operation time in higher
level of parallelism can not account for all of the improvement on
the threshold values; because the crash rate constant per unit time
$\Gamma_{t}$ has been scaled by the amount of time needed to complete
a fault-tolerant QEC step ($\Gamma_{t}=\Gamma_{n}/\tau$), any difference
in $\Gamma_{t}$ is from sources other than difference in $\tau$.
The improvement on the threshold value is due to the effect that when
the level of parallelism is increased, the number of pathways that
generate uncorrectable errors decreases. We emphasize that because
our method of estimating noise threshold does not assume maximal parallelism,
we can access the real threshold value that reflects the limitations
of each individual physical implementations.

\section{Conclusion\label{sec:Conclusion}}

We have applied a noise model based on a generalized effective Hamiltonian
to study the effect of noise on the performance of fault-tolerant
QEC circuits. The model includes realistic physical interactions for
the implementations of quantum gates, and describes the effect of
system-bath interactions by including stochastic fluctuating terms
in the system Hamiltonian. As a result, this method simulates quantum
circuits under physical device conditions, and gives us a full description
of the dissipative dynamics of the quantum computer.

Fault-tolerant QEC circuits implementing either the three qubit bit-flip
code or the five-qubit code were investigated, and the noise threshold
for quantum memory and logical $X$ gate were calculated by comparing
the logical crash rate to the error rate of a bare physical qubit.
The noise threshold of quantum memories using the three qubit bit-flip
code and five qubit code is about $2\times10^{-2}$ and $5\times10^{-4}$,
respectively. The noise threshold of logical $X$ gates using the
three qubit bit-flip code and five qubit code is about $1\times10^{-3}$
and $4\times10^{-5}$, respectively. Note that in our dimensionless
system, these noise strength values should be interpreted as the error
rate per unit time scale $\Delta t=1/\varepsilon$, where $\varepsilon$
is the strength of the control fields. These threshold values are
obtained from an uniform noise model where magnitudes of storage errors
and gate errors are the same. This result indicates that fault-tolerant
quantum computing is possible in systems with strong storage errors.
A possible scenario for such system is the linear nearest-neighbor
architecture, where only nearest-neighbor interactions are available
for two-qubit gates, and excess amount of quantum swap gates have
to be added to the circuit to perform two-qubit gates between qubits
distant in space.

We have also carried out a systematic study on several variables that
can affect the performance of the fault-tolerant QEC procedure for
the three qubit bit-flip code. Our results show that both collective
bath and imperfect projective measurement have minor effects on the
threshold value. However, the repetition protocol and level of parallelism
can significantly change the performance of the fault-tolerant QEC
procedure. Our density matrix results indicate that accepting a syndrome
only when two consecutive syndrome detections agree (protocol B),
which reduces the number of required syndrome detection steps, is
the optimal repetition protocol. Compared to the simple majority vote
algorithm (protocol A), protocol B increases the efficiency of fault-tolerant
QEC at least by a factor of two. Regarding the level of parallelism
in the syndrome detection circuit, in general, a higher level of parallelism
results in a more efficient fault-tolerant QEC circuit. The improvement
can not be fully explained by the shorter operational time for a more
parallelized circuit; we suggest the major contribution for the improvement
comes from the reduction of possible pathways for error propagation.
Since the level of parallelism is actually limited by available physical
resources in reality, it will be interesting to examine and simulate
this factor according to a specific physical implementation of quantum
computer (such as ion-trap or NMR).

Finally, we emphasize that without specifying the specific noise model
and physical device conditions, noise threshold values are of little
usefulness. Our noise model is based on well defined parameters that
reflect realistic device conditions, and provides a full description
for the dissipative dynamics of the quantum computer. As a result,
this noise model enables us to access the real performance of fault-tolerant
QEC for individual physical implementations. We believe that such
information can be useful for the design and optimization of quantum
computers. 

\bibliographystyle{apsrev}
\bibliography{../qcref,../reference,../qmref,../iontrap}

\end{document}